\documentstyle[amssymb,12pt]{article} 
\date{}
\def\frac#1#2{{#1\over#2}} \def\pd#1#2{\frac{\partial#1}{\partial#2}}
\def\matrdos#1#2#3#4{\left(\matrix{#1 & #2 \cr          
                                 #3 & #4 \cr}\right)}

\def\oint{{\relax \int\kern -1. em O}}

\begin{document}

\title{\bf Integrability of Riccati equation
from a group theoretical viewpoint}

\author{Jos\'e F. Cari\~nena and Arturo Ramos}
\maketitle

\begin{center}
Departamento de F\'{\i}sica Te\'orica. Facultad de Ciencias. \\
Universidad de Zaragoza, E-50009, Zaragoza, Spain.
\end{center}  

\begin{abstract}
In this paper  we develop some group theoretical methods 
which are shown to be very useful for a better understanding of 
the properties of the Riccati equation and we discuss some 
of its integrability 
conditions  from a group theoretical perspective. The nonlinear 
superposition principle also arises in a simple way.

\end{abstract}

\def\ba{\begin{eqnarray}}
\def\ea{\end{eqnarray}}
\def\be{\begin{equation}}
\def\ee{\end{equation}}
\def\Eq#1{{\begin{equation} #1 \end{equation}}}
\def\R{\Bbb R}
\def\C{\Bbb C}
\def\Z{\Bbb Z}
\def\a{\alpha}                  
\def\b{\beta}                   
\def\g{\gamma}                  
\def\d{\delta}                  

\def\la#1{\lambda_{#1}}
\def\teet#1#2{\theta [\eta _{#1}] (#2)}
\def\tede#1{\theta [\delta](#1)}
\def\N{{\frak N}}
\def\GR{{\cal G}}
\def\Wei{\wp}
\newcommand{\bea}{\begin{eqnarray}}
\newcommand{\eea}{\end{eqnarray}}

\font\tengoth=eufm10 \font\sevengoth=eufm7 \font\fivegoth=eufm5
\newfam\gothfam
\textfont\gothfam=\tengoth \scriptfont\gothfam=\sevengoth
  \scriptscriptfont\gothfam=\fivegoth
  \def\goth{\fam\gothfam}    
\font\frak=eufm10 scaled\magstep1

\def\bra#1{\langle#1|}
\def\ket#1{|#1\rangle}
\def\goth #1{\hbox{{\frak #1}}}
\def\<#1>{\langle#1\rangle}
\def\cotg{\mathop{\rm cotg}\nolimits}
\def\Map{\mathop{\rm Map}\nolimits}
\def\wt{\widetilde}
\def\const{\hbox{const}}
\def\grad{\mathop{\rm grad}\nolimits}
\def\Div{\mathop{\rm div}\nolimits}
\def\braket#1#2{\langle#1|#2\rangle}
\def\Erf{\mathop{\rm Erf}\nolimits}
\def\matriz#1#2{\left( \begin{array}{#1} #2 \end{array}\right) }
\def\Eq#1{{\begin{equation} #1 \end{equation}}}
\def\deter#1#2{\left| \begin{array}{#1} #2 \end{array}\right| }
\def\pd#1#2{\frac{\partial#1}{\partial#2}}
\def\til{\tilde}

\vskip 1cm

\section{Introduction}

The Riccati equation is very often used in many different fields of 
theoretical physics (see e.g. \cite{CMN}  and
references therein) and this is particularly true during the 
last years,
because 
this differential equation has recently been shown to be related
with the factorization method (see e.g. \cite{CMPR}) and its importance
has been increasing since Witten's introduction of supersymmetric
Quantum Mechanics. On the other side, the first order differential equation
obtained by reduction from linear second order differential equations, 
according to Lie theory, when taking into account the invariance under the  
one-parameter group of dilations, is a Riccati equation. Moreover, from
the mathematical viewpoint it is esentially the only differential 
equation, with one dependent variable, admitting a
non--linear superposition principle, and this fact shows that 
there is a symmetry group for the theory which can be used 
for a better understanding of the properties of such equation.

The main point is that the general solution of a Riccati equation 
cannot be expressed by means 
of quadratures  but in some particular cases. This motivated the study 
of different cases in which such an  explicit solution is possible. 
So, Kamke's book (1959)  presents some integrability conditions which have
also been completed recently by Strelchenya (1991).

The criterion established by  Strelchenya was based on the use of some
transformations by $t$--dependent matrices with values in $SL(2,{\R})$.
We aim in this paper to develop the ideas pointed out by Strelchenya, but
our interpretation will be different. In particular, it will be shown
that his new  criterion 
reduces to a very well known fact: the general solution can be explicitly
written by means of two quadratures if one particular solution of the
differential equation is known.

The paper is organized as follows: In Section 2 we recall 
some well known properties of Riccati equation and we
have summarized its general properties of integrability 
which are known to date. The geometric interpretation of 
the general Riccati equations as a time--dependent vector field in 
the one--point compactification of the real line is given in Section 3.   
The  particular case of Riccati equations with constant coefficients
that is carried out in Section 4 from this geometric approach is
based on the action of the group $SL(2,{\R})$ on the set of such
Riccati equations,
as a way of introducing in a particular case what we will do
 for the general  case  in  Section 5 by means of  the  
group ${\cal G}$ of maps of ${\R}$ in  $SL(2, {\R})$. 
The transformation formulae for 
 the coefficients of the Riccati equation involve 
time derivatives of the coefficients, but we will show that  
these additional terms define a cocycle 
and therefore it is possible to define an affine action of 
such a group ${\cal G}$ on the set of Riccati equations. The 
action so obtained furnishes a method
 for studying the reduction when one or two solutions of Riccati equation 
are known and provides a method of introducing some integrability criteria.
Finally, in the last section the 
non--linear superposition principle 
is revisited from this new group theoretical perspective.

\section{Integrability criteria for Riccati equation}

We recall that the Riccati equation is a non--linear first
order differential equation
\be
\frac{dx(t)}{dt}=a_0(t)+a_1(t)\,x(t)+a_2(t)\,x^2(t)\ 
\label{Riceq}
\ee
and that there is no way of writing the general
solution, in the most general case, by using some quadratures.
However, there are some particular cases for which one can write
 the general solution by such an expression. Of course the
simplest case is when
 $a_2=0$, 
i.e.,  when the equation is linear: then, two quadratures allow us 
to find the 
general solution, given explicitly by 
$$
x(t)=\exp\bigg\{\int_0^ta_1(s)\,ds\,\bigg\} \times \bigg\{x_0+\int_0^t
 a_0(t^{\prime})
        \exp\{ -\int_0^{t^{\prime}} a_1(s) \,ds \} \,dt^{\prime}
\bigg\}\ .
$$

It is also remarkable that under the change of variable 
\be
w=-\frac 1x\label{menosinv}
\ee
the  Riccati  equation (\ref{Riceq}) becomes a new  Riccati equation 
\be
\frac{dw(t)}{dt}=a_0(t)\, w^2(t)-a_1(t)\,w(t)+a_2(t)\ .
\ee
This shows that if in the original
equation  $a_0=0$ (Bernoulli with $n=2$),
then the mentioned change of variable transforms
the given equation into a 
homogeneous linear one, and therefore
the general solution can be written
by means of two quadratures.

Two other integrability conditions can be found in \cite{Kamke}:

a)  The coefficients satisfy $a_0+a_1+a_2=0$.

b) There exist constants $c_1$ and $c_2$ such that
$c_1^2\,a_2+c_1\,c_2\,a_1+c_2^2\,a_0=0$.

In a recent paper Strelchenya (1991) claimed to give a {\sl new\/}
integrability criterion:

c) There exist functions $\alpha(t)$ and $\beta(t)$ such that
\be
a_2+a_1+a_0=\frac d{dt}\log \frac \alpha\beta-\frac{\alpha-\beta}
{\alpha\,\beta}(\alpha \,a_2-\beta\, a_0)\ , \label{strel1}
\ee
which can also  be rewritten as 
\be
\alpha^2\,a_2+ \alpha\,\beta\, a_1+\beta^2\, a_0=\alpha\beta\,
\frac d{dt}\log \frac \alpha\beta\ .\label{strel2}
\ee

All these preceding conditions, including what Strelchenya called
a new integrability condition, are nothing
but three particular cases of a well
known result (see e.g. \cite{HTD}): if one particular solution $x_1$ of 
(\ref{Riceq}) is known,
then the change of variable 
\be
x=x_1+x'\label{trasl}
\ee
leads to a new Riccati equation for which the new coefficient 
$a_0$ vanishes:
\be
\frac{d{x'}}{dt}=(2\,x_1\,a_2+a_1)\,{x'}+a_2\,{x'}^2, \label{Bereq}
\ee
that, as indicated above,  can be reduced to a linear equation with
the change $x'=-1/u$. Consequently, when one particular solution is
known, the general solution can be found by means of two quadratures:
is given by $x=x_1-(1/u)$ with
\bea
u(t)&=&\exp\bigg\{-\int_0^t[2\,x_1(s)\,a_2(s)+a_1(s)]\,ds\,\bigg\}\nonumber\\
    & &\times \bigg\{u_0+\int_0^t a_2(t^{\prime})
        \exp\{ \int_0^{t^{\prime}}[2\,x_1(s)\,a_2(s)+a_1(s)]\,ds \}\,
dt^{\prime}\bigg\}\ .
\label{expresu}
\eea

The criteria a) and b) correspond to the fact that either the constant 
function $x=1$, in case a), or $x=c_1/c_2$, in case b),
are solutions of the given Riccati
equation \cite{Mu 60}. What is not so obvious is that, actually, 
the  condition given in c) is equivalent to say that the function
$x=\alpha/\beta$ is a solution of (\ref{Riceq}).

Moreover, it is also known \cite{HTD} that when not only one but two 
particular solutions of (\ref{Riceq}) are known,
$x_1(t)$ and $x_2(t)$, the general solution can be found
by means of only one
quadrature.
In fact, the change of variable 
\be
\bar{x}=\frac {x-x_1}{x-x_2}\  \label{dossol}
\ee
transforms the original equation into a homogeneous 
linear differential equation in the new variable $\bar{x}$,
$$
\frac{d{\bar{x}}}{dt}=a_2(t)\,(x_1(t)-x_2(t))\,{\bar{x}}\ ,
$$
which has general solution
$$
{\bar{x}}(t)={\bar{x}}(t=0)\,e^{\int_0^t a_2(s)\,(x_1(s)-x_2(s))\,ds}\ .
$$

Alternatively, we can consider the change
\be
x''=(x_1-x_2)\,\frac{x-x_1}{x-x_2}\ ,\label{dossol2}
\ee
and the original Riccati equation~(\ref{Riceq}) becomes
$$
\frac{d{x''}}{dt}=(2\,x_1(t)\,a_2(t)+a_1(t))\,{x''}\ ,
$$
and therefore the general solution can be immediately found:
$$
{x''}={x''}(t=0)\,e^{\int_0^t (2\,x_1(s)\,a_2(s)+a_1(s))\, ds}\ .
$$
We will comment the relation between both changes of variables
and find another possible one later on.

Even more interesting is the following property:
once three particular solutions, $x_1(t),x_2(t),x_3(t)$, are known,
the general solution can be
written, without making use of any quadrature, in the following way:
\be
\frac{x-x_1}{x-x_2}:\frac{x_3-x_1}{x_3-x_2}=k\ ,
\label{superp_formula}
\ee
where $k$ is an arbitrary constant characterizing each particular solution.

Notice that the theorem for uniqueness of solutions of differential
equations shows that the difference between two solutions of the Riccati
equation (\ref{Riceq}) has a constant sign and therefore the difference 
between two different solutions never vanishes, and
the previous quotients are always well defined.

This is a non--linear superposition principle: given three particular
fundamental solutions there exists a superposition function
$\Phi(x_1,x_2,x_3,k)$ 
such that the general solution is expressed as
$x=\Phi(x_1,x_2,x_3,k)$. In this particular case the superposition function
is given by
\be
x=\frac {k\, x_1(x_3-x_2)+x_2(x_1-x_3)}{k\,(x_3-x_2)+(x_1-x_3)}\ .
\ee

This superposition principle and the generalization given by 
the so called Lie--Scheffers theorem \cite{LS}, which has had a revival
 after several interesting papers by Winternitz and coworkers (see e.g. 
\cite{PW}),
have been studied in \cite{CMN} from 
a group theoretical perspective.  We aim here to
investigate other interesting 
properties by using appropriate group theoretical 
methods and the relationship of these properties
with the usual integrability
criteria.

\section {Geometric Interpretation of Riccati equation}

 From the geometric viewpoint the Riccati equation can be considered
as a differential equation determining the integral curves of  the 
time--dependent vector field 
\be
\Gamma=(a_0(t)+a_1(t)x+a_2(t)x^2)\pd{}{x}\ .\label{vfRic}
\ee
 The simplest
case is when all the coefficients $a_i(t)$ are constant, 
because then $\Gamma$, given by (\ref{vfRic}),
is a true vector field. Otherwise, $\Gamma$ is a vector field along 
the projection $\pi:\R\times \R\to \R$, given by $\pi(x,t)=x$
(see e.g. \cite{Car 96} where it is shown that these vector fields
 along $\pi$ also admits integral curves).

The important point is that 
$\Gamma$ is a linear combination with time--dependent coefficients
of the three following vector fields
\be
L_0 =\pd{}{x}\,,        \quad
L_1 =x\,\pd{}{x}\, ,    \quad
L_2 = x^2\,\pd{}{x}\,,  \label{sl2gen}
\ee
that one can check that close on a three-dimensional real Lie  algebra,
with defining relations
\be
[L_0,L_1] = L_0\,,      \quad 
[L_0,L_2] = 2L_1\,,     \quad
[L_1,L_2] = L_2 \,,
\label{conmutL}
\ee
and consequently this Lie  algebra is isomorphic to  $\goth{sl}(2,{\R})$,
which is made up by traceless $2\times 2$ matrices. A basis is given by
\be
M_0=\matriz{cc}{0&1\\0&0}\,, 
\ M_1=\frac{1}{2}\matriz{cc}{1&0\\0&-1}\,,\ M_2=\matriz{cc}{0&0
\\-1&0}\ .
\label{base_matrices}
\ee
Let us remark that the matrices (\ref{base_matrices}) have the 
same commuting relations as the
corresponding $L$'s but with opposite sign, because the identification  
between the $L$'s and the $M$'s is an antihomomorphism of Lie algebras.

Notice also that the commutation relations (\ref{conmutL}) 
show that $L_0$ and $L_1$
generate a two--dimensional Lie subalgebra isomorphic to the Lie algebra
 of the affine group of transformations in one dimension, 
and the same holds for $L_1$ and $L_2$. 

The one--parameter subgroups of local transformations of $\R$ 
generated  by  $L_0$, $L_1$ and  $L_2$ are
$$ 
x\mapsto x+\epsilon\,,\quad x\mapsto e^\epsilon x\,,\quad x
\mapsto\frac x{1-x\epsilon}\ . 
$$

Notice that $L_2$ is not a complete vector field on $\R$. However we can 
do the one-point compactification of $\R$ and then  
$L_0$, $L_1$ and  $L_2$ can be considered as the fundamental vector fields 
corresponding to the action of $SL(2,{\R})$ on the completed real
line $\overline{\R}$, given by 
\begin{eqnarray}
\Phi(A,x)={\frac{\alpha x+\beta}{\gamma x+\delta}},\ \ \ \mbox{if}\
x\neq-{\frac{\delta}{\gamma}},
\nonumber\\
\Phi(A,\infty)={\frac{\alpha}{\gamma}}\ ,\ \ \ \
\Phi(A,-{\frac{\delta}{\gamma}})=\infty,
\nonumber\\
\mbox{when}\ \ \ A=\matrdos {\alpha}{\beta}{\gamma}{\delta}\,\in SL(2,{\R}).
\end{eqnarray}

This suggests us that the group  $SL(2,{\R})$ should play a prominent role
in the study of Riccati equation, as it will be  shown shortly.

\section{The  Riccati equation with constant coefficients}

Let us first consider the simplest case of Riccati equations with constant 
coefficients
\be
\dot x=a_0+a_1\,x+a_2\,x^2 \ .
\label{Ric_coef_ctes}
\ee
As indicated
above, the set of such equations is a $\R$--linear space that 
can be identified with 
the set of fundamental vector fields corresponding to the above
mentioned action  $\Phi$ of  $SL(2,{\R})$ on
the extended real line  $\overline{\R}$. It is very easy to check that
under  the change of variable 
$$x'=\frac {\alpha x+\beta}{\gamma x+\delta}\ ,
\qquad A=\matriz{cc}{\alpha &\beta\\
 \gamma &\delta}\in SL(2,{\R}),\ {\rm {if\ }} x \not =-\frac\delta\gamma\ ,
$$
the original Riccati equation (\ref{Ric_coef_ctes})
becomes a new Riccati equation 
\be
\dot x '=a'_2\,x^{\prime\, 2}+a'_1\,x'+a'_0\ ,
\ee 
where the relation amongst the  old and new coefficients is given by 
\bea
a'_2&=&{\d}^2\,a_2-\d\g\,a_1+{\g}^2\,a_0                   \nonumber\\
a'_1&=&-2\,\b\d\,a_2+(\a\d+\b\g)\,a_1-2\,\a\g\,a_0 \label{camb_ai_ctes}
\qquad.   \\
a'_0&=&{\b}^2\,a_2-\a\b\,a_1+{\a}^2\,a_0   \nonumber               
\eea

This can also be written in a matrix form
\bea
\matriz{ccc}{{a'_2}\\{ a'_1}\\{ a'_0}}
=\matriz{ccc}{{{\d}^2}&{-\d\g}&{{\g}^2}\\
                        {-2\,\b\d}&{\a\d+\b\g}&{-2\,\a\g}\\
        {{\b}^2}&{-\a\b}&{{\a}^2}}\matriz{c}{{ a_2}\\{a_1}\\{a_0}}\ .
        \label{matr_camb_ai_ctes}
\eea

In this way we define a linear  action of the group
 $SL(2,{\R})$ on the set of 
Riccati equations with constant coefficients, i.e., on
the set of fundamental vector fields and therefore 
on the Lie algebra  ${\goth{sl}}(2,{\R})$. The action so obtained is 
nothing but the 
adjoint representation. Moreover, using the Killing--Cartan form on 
${\goth{sl}}(2,{\R})$ we can establish a one-to-one correspondence 
of it  with 
${\goth{sl}}(2,{\R})^*$ and the action turns out to  be the coadjoint 
action. The 
orbits of the action are  then easily found: they are symplectic manifolds
 characterized by the values of the Casimir function corresponding to the 
natural Poisson structure defined on ${\goth{sl}}(2,{\R})^*$, which
is $a_1^2-4\,a_0a_2$.

But the knowledge of the  orbits is very important because in order to solve
the given equation we can analyze whether there exists another easily
solvable Riccati  differential equation in the same orbit or not.

For instance, we would like to find some  constant coefficients 
  $\a,\b,\g,\d$ with $\a\g-\b\d=1$ and such that either
\be
a'_0={\b}^2\,a_2-\a\b\,a_1+{\a}^2\,a_0=0\ ,\label{a0cero}
\ee
or
\be
a'_2={\d}^2\,a_2-\d\g\,a_1+{\g}^2\,a_0=0\ .\label{a2cero}
\ee

 In this case the 
corresponding transformation would reduce the original 
equation to a simpler one.

Notice however that if such coefficients do exist, then 
$x=- {\b}/{\a}=k_1$, or   
$x=- {\d}/{\g}=k_1$, respectively,
is a constant solution of the 
original  Riccati equation. 
Remember that the condition for the existence of
at least one constant solution is  $a_1^2-4\,a_2 a_0\geq 0$.
Conversely, if 
$x=k_1$ is a constant solution then the matrix  
\be
\matriz{cc}{1&-k_1\\0&1}\label{mconst1}
\ee
which corresponds to the change $x'=x-k_1$
will transform the original equation
(\ref{Ric_coef_ctes})
into a new one with 
$a'_0=0$. That is,
\be
\dot {x}'=a'_2\, x^{\prime\, 2}+a'_1\, x'\ ,
\label{Ric_cte_transl}
\ee
where 
$a'_1=2\,k_1a_2+a_1$ and $a'_2=a_2$.

Respectively, a similar assertion follows for the matrix 
\be
\matriz{cc}{1&0\\-k_1^{-1}&1}
\label{mconst2}
\ee
which corresponds to the change
\be
{\til{x}}'=\frac{k_1\,x}{k_1-x}
\label{cambio_const_2}
\ee
which transforms the equation (\ref{Ric_coef_ctes}) into one 
with ${\til a}'_2=0$.

Let us assume now that 
there exists another constant solution  $k_2$, for which a
necessary and sufficient condition is 
$a_1^2-4\,a_0a_2> 0$.
If we do first the change given by (\ref{mconst1}) then $k_2-k_1$ 
will be a solution of (\ref{Ric_cte_transl}) and a new change given 
by the matrix
$$
\matriz{cc}{1&0\\(k_1-k_2)^{-1}&1}\ ,
$$
i.e.,
\be
x''=\frac {(k_1-k_2) x'}{x'+k_1-k_2}\ ,\label{doscons}
\ee
will lead to  a new Riccati equation with 
$a''_2=a''_0=0$, namely 
\be
\dot x''=a''_1 \, x''   \ ,
\label{eq_linhom_cte1}
\ee
with $a''_1=a'_1=2\,k_1\,a_2+a_1$.
If we take into account that if $k_1,\,k_2$ are constant 
solutions of (\ref{Ric_coef_ctes}) the following relations
will be  satisfied
\be
a_1=-a_2\,(k_1+k_2)\ ,\qquad a_0=a_2\,k_1k_2\ ,\label{cardano}
\ee
and then  $a''_1=2\,k_1\,a_2+a_1=a_2\,(k_1-k_2)$.
So, the equation (\ref{eq_linhom_cte1}) 
can be integrated by means of a quadrature: 
$$
x''(t)=x''(0)e^{a_2(k_1-k_2)t}.
$$ 
Notice that the final equation (\ref{eq_linhom_cte1})
is linear, so any constant multiple of the transformation 
(\ref{doscons}) also does the work.

When $a_0\ne 0$ and two constant solutions $k_1,\,k_2$ of 
(\ref{Ric_coef_ctes}) are known
we can also follow an alternative way  because 
the expressions for
$a'_2$ and $a'_0$ in terms of the old coefficients 
are similar, with the interchange of $\beta $ by $\delta$ 
and $\alpha$ by $\gamma$, as can be seen 
in (\ref{camb_ai_ctes}).
Therefore, the knowledge of a constant solution $k_1\ne 0$ can be used
to put first  $\tilde a'_2=0$ 
by means of the transformation (\ref{cambio_const_2})
associated to the  matrix (\ref{mconst2})
obtaining the linear equation
\be
\dot{{\tilde x}'}={\tilde a_1}'\,{\tilde x}'+{\tilde a_0}'
\label{Ric_cte_cotransl}
\ee
where 
$$
{\tilde a}'_0=a_0 , \ \ \  {\tilde a}'_1=a_1+2k_1^{-1}a_0\ .
$$
Then, as another constant solution $k_2$ of (\ref{Ric_coef_ctes})
is known, 
the constant function $k_1\,k_2\,(k_1-k_2)^{-1}$ is a solution of 
(\ref{Ric_cte_cotransl}).
As a consequence, we can consider the transformation
corresponding to the matrix 
\be
\matriz{cc}{1&-k_1k_2(k_1-k_2)^{-1}\\0&1}\ ,
\label{matr2constaltern}
\ee
i.e.,
\be
{\til x}''={\til x}'-\frac{k_1\,k_2}{k_1-k_2}
\label{transf2constaltern}
\ee
which transforms (\ref{Ric_cte_cotransl}) into the
homogeneous linear equation
\be
\dot{{\til x}''}={\til a_1}''\,{\til x}''\ ,
\label{eq_linhom_cte2}
\ee
with
${\tilde a}''_1=a_2\,(k_2-k_1)$.

\section{The general Riccati equation with coefficients depending on 
 $t$}

In the general case in which the coefficients can depend on $t$,
each Riccati equation can be considered as a 
curve in ${\R}^3$, or, in other words, as an element of 
$E=\Map({\R},\,{\R}^3)$.

The point now is that  we can 
transform every function in $\R$, $x(t)$,
under an element of the group
of smooth $SL(2, {\R})$--valued curves 
$\Map({\R},\,SL(2,{\R}))$, which from now on 
will be denoted ${\GR}$, as follows:
\bea
\Theta(A,x(t))={\frac{\a(t) x(t)+\b(t)}{\g(t) x(t)+\d(t)}}\ ,\ \ \ 
\mbox{if\ }\  x(t)\neq-{\frac{\d(t)}{\g(t)}}\ ,
\\
\Theta(A,\infty)={\frac{\a(t)}{\g(t)}}\ ,\ \ \ \ \Theta(A,-{\frac{\d(t)}{\g(t)}})=\infty\ ,\\ \mbox{when}\ 
A=\matriz{cc} {{\a(t)}&{\b(t)}\\{\g(t)}&{\d(t)}}\,\in\Map({\R},\,SL(2,{\R}))\ ,
\label{Agauge}
\eea
and it is now easy to check that the Riccati equation (\ref{Riceq})
transforms under these changes into a new 
 Riccati equation with coefficients given by 
\bea
a'_2&=&{\d}^2\,a_2-\d\g\,a_1+{\g}^2\,a_0+\g {\dot{\d}}-\d \dot{\g}\ ,
\label{ta2}\\
a'_1&=&-2\,\b\d\,a_2+(\a\d+\b\g)\,a_1-2\,\a\g\,a_0   
       +\d \dot{\a}-\a \dot{\d}+\b \dot{\g}-\g \dot{\b}\ ,  \label{ta1}  \\
a'_0&=&{\b}^2\,a_2-\a\b\,a_1+{\a}^2\,a_0+\a\dot{\b}-\b\dot{\a} \ .
 \label{ta0} 
\eea

Some particular instances of transformations
of this type are those given by
(\ref{menosinv}), (\ref{trasl}), (\ref{dossol}) and (\ref{dossol2}).

We can use this expression for defining an affine action of the group 
 ${\GR}$ on the set of general
Riccati equations. The relation amongst new and old coefficients
can be written in a matrix form
\bea
\matriz{c}{a'_2\\a'_1\\a'_0} 
&=&
\matriz{ccc}
{{\d}^2&{-\d\g}&{{\g}^2} \\ {-2\,
\b\d}&{\a\d+\b\g}&{-2\,\a\g}\\{{\b}^2}&{-\a\b}&{{\a}^2}}
\matriz{c}{{a_2}\\
{a_1}\\{a_0}}
\nonumber\\     
& &+\matriz{c}{\g {\dot{\d}}-\d \dot{\g}\\
\d \dot{\a}-\a \dot{\d}+\b \dot{\g}-\g \dot{\b}\\
 \a \dot{\b}-\b \dot{\a}}\ .
\label{transf_ai_dept_matricial}
\eea

In the first term of the right hand side we see the adjoint action. As 
far as the second term is concerned, we can check that it is a 1-cocycle
for the adjoint action 
because if $A$ is given by (\ref{Agauge}), then 
$$
\dot A=\matriz{cc}{\dot{\a}&\dot{\b}\\ \dot{\g}&\dot{\d}}\ \ \
{\rm and }\ \ \ A^{-1}=\matriz{cc}{\d&-\b\\-\g&\a}\ ,
$$
and  therefore
$$
\theta(A)=\dot A A^{-1}
=\matriz{cc}{\d\dot{\a}-\g\dot{\b}&\a\dot{\b}-\b\dot{\a}\\
\d\dot{\g}-\g\dot{\d}&\a\dot{\d}-\b\dot{\g}}\ ,
$$
is a zero trace matrix because of the condition $\a\d-\b\g=1$. Then,
we can make use of the natural identification of the 
Lie algebra ${\goth{sl}}(2,{\R})$
with the one of zero trace matrices,
taking into account our election of the basis (\ref{base_matrices}).
We arrive to the following value for the image of $\theta(A)$ under  
such identification, which with a slight abuse of notation we also 
denote by $\theta(A)$:  
$$
\theta(A)=\matriz{c}{\g \dot{\d}-\d \dot{\g}\\
\d \dot{\a}-\a \dot{\d}+\b \dot{\g}-\g \dot{\b}\\
                \a \dot{\b}-\b \dot{\a} }\ ,
$$
that is, the second term of the right side 
of (\ref{transf_ai_dept_matricial}).
It is quite simple to check that the cocycle condition holds, because 
\bea
\theta(A_2A_1)&=&(A_2A_1)\,{\dot{}}\,(A_2A_1)^{-1}
                        =(\dot A_2A_1+A_2\dot A_1)A_1^{-1}A_2^{-1} \nonumber\\
              &=& \dot A_2A_2^{-1}+(A_2\dot A_1)A_1^{-1}A_2^{-1}\ ,
\eea
or in a different way,
$$\theta(A_2A_1)=\theta(A_2)+A_2\theta(A_1)A_2^{-1}\ ,
$$
which is the 1--cocycle condition for the adjoint action. 
Consequently, see e.g \cite{LM87}, 
the expression (\ref{transf_ai_dept_matricial}) defines an affine 
action of ${\GR}$ on the set of general Riccati equations.
In other terms, to transform the coefficients of a general Riccati 
equation by means of two successive transformations of type 
(\ref{transf_ai_dept_matricial}) which are associated respectively with
two elements $A_1,\,A_2$ of ${\GR}$, gives exactly
the same result as doing only {\sl one\/} transformation of type
(\ref{transf_ai_dept_matricial}) with associated element $A_2\,A_1$ of
${\GR}$.

In similar way to what happens in the constant coefficients case, we can
 take advantage of some particular transformation to reduce a given 
equation to a simpler one. So, (\ref{ta1}) shows that if we choose
 $\beta =\gamma=0$ and $\delta=\alpha^{-1}$, i.e., 
\be
\matriz{cc}{\alpha&0\\0&\alpha^{-1}} \ ,
\label{prop}
\ee
then $a'_1=0$ if and only if the function $\alpha $ is such that 
$$a_1=-2\,\frac{\dot \a}\a\ ,
$$
equation which has the particular solution 
$$
\a=\exp\left[-\frac 12\int a_1(t) dt \right]\ ,
$$
i.e., the change is $ x'=e^{-\phi}x$ with $\phi=\int a_1(t)\, dt$,
and then $a'_2=a_2e^{\phi} $ and $a'_0=a_0e^{-\phi}$, 
which is the property {\bf{3-1-3.a.i}} of \cite{Mu 60}. 
In fact, under the transformation (\ref{prop})
\be
a'_2=\alpha^{-2} a_2\ ,\quad
a'_1=a_1+2\, \frac {\dot \alpha}\alpha\ ,\quad 
a'_0=\alpha^{2} a_0\ , \label{deralfa}
\ee
and therefore with the above choice for  $\alpha$ we see that 
$a'_1=0$.

If we use instead $\alpha=\delta=1$, $\gamma=0$, 
the function $\beta$ can be 
chosen in such a way that $a'_1=0$ if and only if
$$\beta=\frac{a_1}{2a_2}\ ,
$$
and then 
$$a'_0=a_0+\dot{\beta}-\frac{a_1^2}{4a_2}\ ,\ a'_2=a_2\ ,
$$
which is the property  {\bf{3-1-3.a.ii}} of \cite{Mu 60}.

As another  instance, the original equation (\ref{Riceq})
would be reduced to one with $a'_0=0$ if and only if
there exist functions $\alpha(t)$ and  $\beta(t)$ such that
$$
{\b}^2 a_2-\a \b a_1+{\a}^2 a_0+\a \dot{\b}-\b \dot{\a}=0\ .
$$
This was considered in \cite{Stre}, although written in the
slightly modified way (\ref{strel1}),
as  a criterion for the integrability 
of Riccati equation. 

We should remark as an important fact that when dividing the preceding 
expression by 
 $\alpha^2$ we find that 
$x_1=-\beta/\alpha$ is a solution of the original Riccati equation, and
conversely,
if a  particular solution is known, $x_1$, then the element of ${\GR}$ 
\be
\matriz{cc}{1&-x_1\\0&1}
\label{mvar1}
\ee
with associated change 
\be
x'=x-x_1\ ,
\label{cvar1}
\ee
will transform the equation (\ref{Riceq}) into a new one with
$a'_0=0$, $a'_2=a_2$ and $a'_1=2\,x_1\,a_2+a_1$,
i.e., equation (\ref{Bereq}),
 which can be easily  integrated by two quadratures.
Consequently, the criterion given in \cite{Stre} is nothing but the well 
known fact that once a particular solution is known, the original Riccati
 equation can be reduced to a Bernoulli one and therefore the general 
solution can be easily found. However, in our opinion the previous
 remark gives a very appropriate  group theoretical 
explanation of the convenience of the change of 
variables given by (\ref{trasl}).

As a particular case of the previous, one can interpret the property
{\bf{3-1-3.b.iii}} of \cite{Mu 60}, which assumes that a special solution
$x_1$ of (\ref{Riceq}) is known such that the quantity 
\be
X(t)=2\,x_1(t)\,a_2(t)+a_1(t)
\label{Xmurph}
\ee
is determined by the equation
\be
\dot{x}_1=-a_2\,x_1^2+X(t)\,x_1+a_0,
\label{propmurph}
\ee
and then considers three possible values of $X(t)$.
Replacing~(\ref{Xmurph}) in~(\ref{propmurph}) we 
simply recover the fact that $x_1$ is a solution of (\ref{Riceq}). 
The quantity $X(t)$ is nothing but
the coefficient $a'_1$ obtained with the transformation given by
the matrix~(\ref{mvar1}). Selecting $X(t)$ to be $0$, $-\dot a_2/a_2$ 
and $a_1-2\sqrt{a_0\,a_2}$, respectively, means to consider special types 
of the Riccati equation with one {\sl known} particular solution which 
can be expressed in terms of the coefficients of the equation,
 $a_i(t)$, and their time derivatives, $\dot a_i(t)$, and 
whose general solution involves an immediate 
quadrature, as it can be seen in~(\ref{expresu}). The special case 
which is said in~\cite{Stre} to be absent in~\cite{Kamke} is simply
expressed in these terms by saying that 
$X(t)=a_0-a_2+\dot{a_0}/a_0-\dot{a_2}/a_2$.
We would like to remark that the properties {\bf{3-1-3.a.i}}, 
{\bf{3-1-3.a.ii}} and {\bf{3-1-3.b.iii}} of \cite{Mu 60} can also
be found in \cite{Kamke}.

As indicated in the constant case, we can also follow a similar path
by first reducing the 
original equation (\ref{Riceq}) to a new one with ${\tilde{a}'_2}=0$.
Then, we should look for functions
$\g(t)$ and $\d(t)$ such that 
$$
{\tilde{a}'_2}={\d}^2 a_2-\d \g a_1+{\g}^2 a_0
+\g {\dot{\d}}-\d \dot{\g}=0\ .
$$

This equation is similar to the one satisfied by $\a$ and $\b$ 
in order to obtain $a'_0=0$ with the replacement of $\b$ by $\d$
and $\a$ by $\g$, and therefore we should consider the transformation given 
by the  element of ${\cal G}$ 
\be
\matriz{cc}{1&0\\-x_1^{-1}&1}\ ,
\label{mvar2}
\ee
that is,
\be
\til{x}'=\frac{x_1\,x}{x_1-x}
\label{cvar2}
\ee
in order to obtain a new Riccati with ${\tilde{a}'_2}=0$.
More explicitly, the new coefficients are 
\be
\til{a}'_2=0\,,\quad             
\til{a}'_1=\frac{2\,a_0}{x_1}+a_1\,,\quad        
\til{a}'_0=a_0\ ,        
\ee
i.e., the original Riccati equation (\ref{Riceq}) becomes
\be
\frac{d\tilde x'}{dt}=\left(\frac{2\,a_0}{x_1}+a_1\right)\tilde x'+
a_0\ .
\label{seg}
\ee

Therefore, the transformation (\ref{cvar2})
will {\sl directly} produce a linear equation (\ref{seg}). Such
a change seems to be absent in the literature as far as we know.

Let us suppose now that 
another solution $x_2$ of (\ref{Riceq}) is also known.
If we make the change (\ref{cvar1}) the difference  
$x_2-x_1$ will be a solution of the resulting equation
(\ref{Bereq}) and 
therefore, after using  the change given 
by (\ref{mvar1}), the element of ${\GR}$  
\be
\matriz{cc}{1&0\\(x_1-x_2)^{-1}&1}
\label{mvar1b}
\ee
will transform the Riccati equation (\ref{Bereq}) into a new one
with $a''_2=a''_0=0$ and $a''_1=a'_1=2\,x_1\,a_2+a_1$, namely,
\be
\frac{dx''}{dt}=(2\,x_1\,a_2+a_1)\,x'' \ , 
\label{dossol1}
\ee 
which can be integrated with just one quadrature. 
This fact can also be considered 
as a very appropriate  group theoretical explanation of the introduction 
of the change of variable (\ref{dossol2}).

In fact, we can check directly that if we use the transformation with 
$\alpha=1,\,\beta=0,\, \delta=1$ and
$\gamma=(x_1-x_2)^{-1}$ over the coefficients of (\ref{Bereq}), 
then we find  that still $a''_0=0$ 
and
$$a''_2=a_2-(x_1-x_2)^{-1} a'_1 +(x_1-x_2)^{-2}
(\dot x_1-\dot x_2)\ ,
$$
and as $x_1$ and $x_2$ are solutions of (\ref{Riceq}), taking the
difference we see that
\be
 \dot x_1-\dot x_2=a_1(x_1-x_2)+a_2(x_1^2-x_2^2)\ ,\label{expdif}
\ee
from which we obtain
\bea
a''_2&=&(x_1-x_2)^{-2}\{a_2(x_1-x_2)^{2}+(x_2-x_1)(a_1+2x_1a_2)
                                                        \nonumber\\
& &+a_1(x_1-x_2)+a_2(x_1^2-x_2^2)\}=0\ .                \nonumber
\eea

The composition of both transformations (\ref{mvar1}) and 
(\ref{mvar1b}) leads to the element of ${\GR}$
\be
\matriz{cc}{1 &-x_1\\
(x_1-x_2)^{-1}&-x_2(x_1-x_2)^{-1}}
\label{comp_dos_sols}
\ee
and therefore to the transformation (\ref{dossol2}). 

Now we can compare the transformations (\ref{dossol}) and (\ref{dossol2}).
The first one corresponds to the element of ${\GR}$ 
(we assume that $x_1(t)>x_2(t)$, for all  $t$)
\be
\frac 1{\sqrt{x_1-x_2}}\matriz{cc}{1 &-x_1\\
1&-x_2}\ ,
\label{dos_sol_stand}
\ee
and therefore both matrices (\ref{comp_dos_sols}) 
and (\ref{dos_sol_stand})
are obtained one from the other 
by multiplication by
an element of type
(\ref{prop}) with $\alpha=(x_1-x_2)^{-1/2}$, 
and then (\ref{deralfa}) relates
the coefficients $a''_1$ and $\bar a_1$
arising after one or the other transformation when taking into 
account (\ref{expdif}):
$$\bar a_1=a''_1-a_1-a_2(x_1+x_2)=a_2(x_1-x_2)\ ,
$$
as expected.

By the other hand,
If we use first the change of variable given by (\ref{cvar2}),
the function  
$$
\tilde x'_2=\frac{x_2\,x_1}{x_1-x_2}
$$
will be a solution of (\ref{seg}). Then, a new transformation given by
the element of ${\GR}$
\be
\matriz{cc}{1&-\frac{x_2\,x_1}{x_1-x_2}\\{0}&{1}}
\label{mvar2b}
\ee
will lead to a new equation in which 
${\til{a}}''_0=0$. More explicitly,
\be
{\til{a}}''_2=0\ ,\quad       
{\til{a}}''_1=\til{a}'_1=\frac{2\,a_0}{x_1}+a_1\ ,\quad
{\til{a}}''_0=0\ .                              
\ee
The composition of the two transformations is 
\be
\matriz{cc}{1&-\frac{x_2\,x_1}{x_1-x_2}\\{0}&{1}}
\matriz{cc}{1&0\\-\frac{1}{x_1}&{1}}=
\matriz{cc}{\frac{x_1}{x_1-x_2}&-\frac{x_2\,x_1}{x_1-x_2}\\
-\frac{1}{x_1}&{1}}\ ,
\ee
which corresponds to the change of variable 
\be
{\til x}''=\frac{x_1^2}{(x_2-x_1)}\frac{(x-x_2)}{(x-x_1)}\ ,
\label{ultima}
\ee
leading to the homogeneous linear equation
\be
\dot {{\til x}}''=\bigg(\frac{2\,a_0}{x_1}+a_1\bigg)
{\til x}''\ ,\label{eqult}
\ee
which can be integrated by means of just one quadrature.

\section{The non--linear superposition principle}

In a recent paper \cite{CMN} the non--linear superposition principle for
the Riccati equation was considered by using the Wei--Norman method of
integrating differential equations on a Lie group, the superposition
principle for Riccati equation arising in a very natural way.

The most important fact is that we can obtain the general solution even 
without 
solving the differential equation  in the group. This is the case when we
 know a fundamental set of solutions that can be used to determine the 
explicit time--dependence of the canonical coordinates: this is the reason for 
the existence of such a superposition principle (for more details,
 see \cite{CMN}).

This superposition principle can also be understood from a group
 theoretical viewpoint. 
Let us now suppose that we know three particular solutions 
$x_1,\,x_2,\,x_3$ of (\ref{Riceq})
 and we can assume that $x_1>x_2>x_3$ for any value of the parameter $t$.  
Following the method described in the previous section we can 
use the two first solutions for reducing the Riccati equation 
to the simpler form 
of a linear equation, either to
\be
\dot x''= (2\,x_1\,a_2+a_1)\, x\ ,
\label{lineal_final_1}
\ee
or
\be
\dot {{\til x}}''=\bigg(\frac{2\,a_0}{x_1}+a_1\bigg)
{{\til x}}''\ .
\label{lineal_final_2}
\ee

The set of solutions of such differential equations 
is an one-dimensional
linear space, so it suffices to know a particular solution 
to find the general solution. As we know that 
\be 
x''_3=(x_1-x_2)\,\frac{x_3-x_1}{x_3-x_2}
\ee 
is then a solution of equation (\ref{lineal_final_1}), and
\be
{\til x}''_3=\frac{x_1^2}{(x_2-x_1)}
\frac{(x_3-x_2)}{(x_3-x_1)}
\ee
is a solution of (\ref{lineal_final_2}), 
we can take  advantage of an appropriate  diagonal 
element of ${\GR}$ of the form  
$$
\matriz{cc}{z^{-1/2}&0\\0&z^{1/2}}\ ,
$$
with $z$ being one of the two mentioned solutions 
in order to reduce  
the equations either to  
${\dot x}'''=0$ or  
$\dot{\tilde{x}}{}'''=0$, respectively. These last 
equations have the general solutions  
$$
x'''=k \  ,
$$
or 
$$
{\tilde x}'''=k \  ,
$$
which show the superposition formula (\ref{superp_formula}). 

More explicitly, for the first case (\ref{lineal_final_1}) 
the product transformation will be given by 
\Eq{\matriz{cc}{\sqrt{\displaystyle{\frac{(x_2-x_3)}
{(x_1-x_3)(x_1-x_2)}}}&-x_1
\sqrt{\displaystyle{\frac{(x_2-x_3)}{(x_1-x_3)(x_1-x_2)}}}\\
\displaystyle{\sqrt{\frac{(x_1-x_3)}{(x_2-x_3)(x_1-x_2)}}}&-x_2
\displaystyle{\sqrt{\frac{(x_1-x_3)}{(x_2-x_3)(x_1-x_2)}}}}\ ,}
or written in a different way
\be
\frac {-1}{\sqrt{(x_1-x_2)(x_1-x_3)(x_2-x_3)}}
\matriz{cc}{x_2-x_3 &-x_1(x_2-x_3)\\
x_1-x_3 &-x_2(x_1-x_3)}\ .
\ee   
 The transformation defined by this element of ${\GR}$ is 
\be
x'''=\frac {(x-x_1)(x_2-x_3)}{(x-x_2)(x_1-x_3)}
\ee
and therefore we arrive in this way to the superposition
principle giving the general
solution of the Riccati equation (\ref{Riceq}) in  terms of  
three particular solutions and the value of a 
constant $k$ characterizing each particular solution:
\be
\frac {(x-x_1)(x_2-x_3)}{(x-x_2)(x_1-x_3)}=k\ .
\ee   
The other case (\ref{lineal_final_2}) can be treated in 
a similar way, leading also to the no linear superposition
principle of the Riccati equation.

\section{Conclusions and outlook}

We have carried out the application of group theoretical 
methods in order to understand several well known properties
of Riccati equation that
 seems to have been unrelated until now. This work has been 
motivated 
by the recent applications in Supersymmetric Quantum Mechanics 
and the factorization problems, on the physical side, and 
from the mathematical viewpoint 
by a very interesting property of such equation: it admits
 a non--linear superposition principle.  This property was
 shown by Lie to be related with the particular form of the Riccati
equation \cite{LS}, because the time--dependent vector field of the general Riccati 
 equation can be written as a linear combination with
time--dependent  coefficients of vector fields closing on 
a ${\goth{sl}}(2,{\R})$ algebra. 

The use of such group theoretical methods has allowed us to 
get a better understanding of the reduction of the problem when 
some particular solutions are known, and we have discovered
 a previously,  as far as we know,  unknown
alternative way for such reduction, that is, the one given
by the changes (\ref{mvar2}) and (\ref{mvar2b}). In particular, 
 we have obtained explicitly the non--linear 
superposition principle of the 
Riccati equation.
As a by-product, we have made clear the meaning of the
integrability
conditions pointed out by Strelchenya in  \cite{Stre}
as well as other previously known criteria.

The question is now open to the possible generalization of 
these properties in the case of other differential equation
systems admitting a non--linear superposition principle, for
which similar techniques should be useful. 

For such a generalization, it might be useful the observation that
for some particular cases, the studied  Bernoulli
and linear ones, for which the general solution is easily found, 
the time--dependent vector field associated to the
equation takes values in a Lie subalgebra isomorphic to the Lie algebra 
of the  affine group of transformations in one dimension.  
Those Riccati equations 
that are in the same orbit under the action of $\cal G$ as one of these
particular cases can be reduced to them and therefore they are 
integrable. 

The elements of ${\GR}$ which take the Riccati equation to 
their reduced form provided some particular solutions are known, are 
constructed {\sl with} such solutions, giving an explanation 
to the previously known properties of the Riccati equation
involving changes of variable which seemed to be 
rather miraculous.

We hope that the
application of these techniques will also be enlightening for the 
above mentioned problems of factorization of a quantum
Hamiltonian as a product $A^{\dag} A$ of first order differential
operators and the corresponding problem of related 
Hamiltonian operators \cite{CMPR}.

\bigskip

{\parindent 0cm{\Large \bf Acknowledgements.}} 

\medskip
One of the authors (A.R.) thanks Spanish Ministerio de 
Educaci\'on y Cultura for a FPI grant, research project 
PB96--0717. Support of Spanish
DGES (PB96--0717) is also acknowledged. A.R. would also 
thank F. Cantrjin and W. Sarlet for their warm 
hospitality and fruitful discussions at the University
of Ghent (Belgium), where part of this work was done.
We would also thank G. Gaeta and
M.A. Rodr\'{\i}guez for  critical
reading of the manuscript.


\begin{thebibliography}{xx}

\bibitem[1]
{CMN}  Cari\~nena J.F., Marmo G. and Nasarre J.,
{\em The non--linear superposition principle and the Wei--Norman method}, 
Int. J. Mod. Phys. {\bf 13}, 3601--27 (1998).

\bibitem[2]
{CMPR} Cari\~nena J.F., Marmo G., Perelomov A.M. and Ra\~nada M.F.,
{\em Related operators  and exact solutions  of Schr\"odinger equations},
Int. J. Mod. Phys.{\bf}, (to appear) (1998).

\bibitem[3]
{Kamke}  Kamke E., 
{\sl Differentialgleichungen: L\"osungsmethoden und L\"osungen}, 
Akademische Verlagsgeselischaft, Leipzig (1959).

\bibitem[4]
{HTD} Davis H.T.,
 {\sl  Introduction to Nonlinear Differential and Integral Equations},
Dover, New York (1962).

\bibitem[5]
{Mu 60}   Murphy G.M.,
{\sl Ordinary Differential equations and their solutions}, 
Van Nostrand, New York (1960).

\bibitem[6]
{LS}   Lie S. and Scheffers G.,
{\sl Vorlesungen \"uber continuierlichen Gruppen mit geometrischen und anderen Anwendungen},
  Teubner, Leipzig (1893).

 \bibitem[7]
{PW}  Winternitz P.,
{\it Lie groups and solutions of nonlinear differential equations},
in Nonlinear Phenomena,  K.B. Wolf Ed., Lecture Notes in Physics {\bf 189},
Springer-Verlag N.Y (1983).

\bibitem[8]
{Car 96} Cari\~nena J.F., {\em Sections along maps in Geometry and Physics},
Rend. Sem.  Mat. Univ. Pol.  Torino {\bf 54},  245--56  
 (1996).   

\bibitem[9]
{LM87} Libermann P.  and  Marle Ch.-M.,
{\sl   Symplectic Geometry and Analytical Mechanics},  
Reidel, Dordrecht (1987).

\bibitem[10]
{Stre} Strelchenya V.M.,
{\em A new case of integrability of the general Riccati equation and its
application to relaxation problems},
J. Phys. A: Math. Gen. {\bf 24}, 4965--67 (1991).
\end{thebibliography}
\end{document}